\documentclass[11pt]{article}

\begin{document}

\newcommand{\half}{\frac{1}{2}}
\newcommand{\inv}{^{-1}}

\title{Topological Charge of ADHM Instanton on $\mathbf{R}^2_{\mathrm{NC}}\times\mathbf{R}^2$}
\author{Yu Tian \\{\it Institute of Theoretical Physics, Chinese Academy of Sciences}\\
{\it P. O. Box 2735, Beijing 100080, China}\\
{\tt ytian@itp.ac.cn}}

\maketitle

\begin{abstract}
We have calculated the topological charge of $U(N)$ instantons on
non-degenerate noncommutative space time to be exactly the
instanton number $k$ in a previous paper \cite{Tian3}. This paper,
which deals with the degenerate
$\mathbf{R}^2_{\mathrm{NC}}\times\mathbf{R}^2$ case, is the
continuation of \cite{Tian3}. We find that the same conclusion
holds in this case, thus complete the answer to the problem of
topological charge of noncommutative $U(N)$ instantons.
\end{abstract}

\newpage

In a previous article \cite{Tian3}, we have explicitly shown that
the topological charge of $U(N)$ instantons on non-degenerate
noncommutative space time\footnote{For noncommutative geometry and
field theory, please see \cite{Connes,K S,D N,Szabo}.}
$\mathbf{R}^4_{\mathrm{NC}}$ is equal to the instanton number $k$,
which appears in the noncommutative ADHM construction \cite{N S}
as the dimension of the vector space $V$. There we left the
problem of degenerate noncommutative parameter $\theta$. We
mentioned that our method ceased to work in the limit of
$\theta_2\rightarrow 0$ because of the loss of a Fock space
representation for $z_2$ and $\bar{z}_2$.\footnote{Here we follow
the notation of \cite{Tian3}.}

In this degenerate limit, the noncommutative space time becomes a
direct product $\mathbf{R}^2_{\mathrm{NC}}\times\mathbf{R}^2$. It
is well known that field theories on $\mathbf{R}^4_{\mathrm{NC}}$
suffer from a common problem of non-unitarity. But the problem
does not happen on $\mathbf{R}^2_{\mathrm{NC}}\times\mathbf{R}^2$
if we require a time in the commutative part. For this reason,
$\mathbf{R}^2_{\mathrm{NC}}\times\mathbf{R}^2$ may be of more
physical relevance than $\mathbf{R}^4_{\mathrm{NC}}$. Literature
has also shown that nonsingular ADHM instanton solutions can be
constructed on $\mathbf{R}^2_{\mathrm{NC}}\times\mathbf{R}^2$
without involving any projected states \cite{CKT,KLY}.

In the following paragraphs, we will analytically calculate the
topological charge of $U(N)$ instantons on
$\mathbf{R}^2_{\mathrm{NC}}\times\mathbf{R}^2$. As the
continuation of \cite{Tian3}, this paper will be brief about the
relevant background and rely on some results in \cite{Tian3}.

First let us recall the commutator of complex coordinates:
\begin{equation}\label{z commutator}
[\bar{z}_{1},z_{1}]=\theta_{1}=\zeta,\quad
[\bar{z}_{2},z_{2}]=\theta_{2}=0.
\end{equation}
Now we introduce a Fock space representation for $z_1$ and
$\bar{z}_1$:
\begin{eqnarray}
z_{1}|n\rangle&=&\sqrt{\zeta}\sqrt{n+1}|n+1\rangle,\\
\bar{z}_{1}|n\rangle&=&\sqrt{\zeta}\sqrt{n}|n-1\rangle.
\end{eqnarray}
Then the integration on
$\mathbf{R}^2_{\mathrm{NC}}\times\mathbf{R}^2$ can be expressed as
\begin{equation}
\int d^4\!x\,\mathcal{O}(x)=\zeta\pi\sum_{n=0}^\infty\int
d^2\!x\,\langle n|\mathcal{O}(x)|n\rangle,
\end{equation}
where $d^2\!x=dx^3 dx^4$ \cite{KLY}.

Using the Corrigan's identity \cite{CGOT}, we have the following
expression of topological charge of (ASD) ADHM instanton on
$\mathbf{R}^2_{\mathrm{NC}}\times\mathbf{R}^2$ with instanton
number $k$ \cite{Tian3}:
\begin{eqnarray}
Q&=&\frac{1}{16\pi^{2}}\int d^4\!x\,\mathrm{Tr}_{N}(F_{mn}F_{mn})\\
&=&\frac{1}{16\pi^{2}}\int
d^4\!x\,(8\partial\bar{\partial}+2\partial_{\mu}\partial_{\mu})\mathrm{Tr}[b^{\dagger}(2-\Delta
f\Delta^{\dagger})b f]\\
&=&\frac{\zeta}{\pi}\sum_{n=0}^\infty\int d^2\!x\,\langle
n|(\half\partial\bar{\partial}+\frac{1}{8}\partial_{\mu}\partial_{\mu})\mathrm{Tr}[(2-\Delta'f\Delta'^{\dagger})f]|n\rangle,
\end{eqnarray}
where $\partial\equiv\partial_{z_1}$,
$\bar{\partial}\equiv\partial_{\bar{z}_1}$, $\mu=3,4$, and
$\Delta'\equiv b^{\dagger}\Delta$ \cite{Tian2}.

Similarly as in \cite{Tian3}, we have the truncated summation
\begin{equation}\label{sum}
\sum\limits_{n=0}^{N}\langle
n|\partial\bar{\partial}\mathcal{O}(x)|n\rangle=\frac{N+1}{\zeta}(\langle
N+1|\mathcal{O}|N+1\rangle-\langle N|\mathcal{O}|N\rangle),
\end{equation}
using the Stokes-like theorem. Let
\begin{equation}
\mathcal{O}=(2-\Delta'f\Delta'^{\dagger})f
\end{equation}
again as in \cite{Tian3}, we have for large $N$
\begin{equation}
\langle N|\mathcal{O}|N\rangle=(\zeta N+x^{\mu}x^{\mu})\inv+(\zeta
N+x^{\mu}x^{\mu})^{-2}C+O(N^{-3}),
\end{equation}
where $C$ is some constant $2k\times 2k$ matrix.\footnote{The
non-singularity of each term for arbitrary $x^{3,4}$ in the
summation (\ref{sum}) can be analyzed by the same method as in the
appendix of \cite{Tian3}.} So
\begin{equation}
\langle N+1|\mathcal{O}|N+1\rangle-\langle N|\mathcal{O}|N\rangle
=\frac{-\zeta}{(\zeta N+x^{\mu}x^{\mu})[\zeta
(N+1)+x^{\mu}x^{\mu}]}+O(N^{-3}).
\end{equation}
Thus we obtain one part of the topological charge
\begin{eqnarray}
Q_1&=&\frac{\zeta}{2\pi}\lim_{N\rightarrow\infty}\mathrm{Tr}\int
d^2\!x\,\sum_{n=0}^N\langle
n|\partial\bar{\partial}[(2-\Delta'f\Delta'^{\dagger})f]|n\rangle\\
&=&\frac{-\zeta}{2\pi}\lim_{N\rightarrow\infty}\mathrm{Tr}[N\int
d^2\!x\,(\zeta N+x^{\mu }x^{\mu })^{-2}+O(N^{-2})]\\
&=&-k.
\end{eqnarray}

Noting that $\mathcal{O}$ is an ordinary function (matrix) in
$x^{3,4}$ and a bounded operator \cite{Tian3} on the Fock space,
we immediately work out the other part of the topological charge
\begin{equation}
Q_2=\frac{\zeta}{8\pi}\mathrm{Tr}\sum_{n=0}^\infty\langle n|\int
d^2\!x\,\partial_{\mu}\partial_{\mu}[(2-\Delta'f\Delta'^\dagger)f]|n\rangle=0,
\end{equation}
where we have used the Stokes theorem and the following asymptotic
as $|x^{3,4}|\rightarrow\infty$:
\begin{equation}
f\rightarrow\frac{1}{x^\mu x^\mu}.
\end{equation}

In conclusion, we obtain the topological charge
\begin{equation}
Q=Q_1+Q_2=-k,
\end{equation}
which is expected for an ASD instanton configuration. For an SD
instanton on $\mathbf{R}^2_{\mathrm{NC}}\times\mathbf{R}^2$ with
instanton number $k$, we can similarly have
\begin{equation}
Q=k,
\end{equation}
which completes the answer to the problem of topological charge of
noncommutative $U(N)$ instantons.

\section*{Acknowledgments}

I would like to thank Dr. Jian Dai, Prof. Chuan-Jie Zhu and Prof.
Xing-Chang Song for helpful discussions.


\begin{thebibliography}{99}

\bibitem{Tian3} Y. Tian, C.-J. Zhu and X.-C. Song, {\it Topological Charge of Noncommutative ADHM Instanton}, Mod. Phys. Lett. A {\bf 18} (2003) 1691-1703 [hep-th/0211225].

\bibitem{Connes} A. Connes, {\it Noncommutative Geometry}, Academic Press, New York, 1994.

\bibitem{K S} A. Konechny and A. Schwarz, \textit{Introduction to
M(atrix) Theory and Noncommutative Geometry}, Phys. Rept. {\bf
360} (2002) 353-465 [hep-th/0012145] [hep-th/0107251].

\bibitem{D N} M. R. Douglas and N. A. Nekrasov, \textit{Noncommutative
Field Theory}, Rev. Mod. Phys. 73 (2002) 977, hep-th/0106048.

\bibitem{Szabo} R. J. Szabo, \textit{Quantum Field Theory on
Noncommutative Spaces}, Phys. Rept. \textbf{378} (2003) 207-299
[hep-th/0109162].

\bibitem{N S} N. Nekrasov and A. Schwarz, \textit{Instantons on
Noncommutative $R^{4}$, and (2,0) Superconformal Six Dimensional
Theory}, Commun. Math. Phys. 198 (1998) 689, hep-th/9802068.

\bibitem{CKT} C.-S. Chu, V. V. Khoze and G. Travaglini, {\it
Notes on Noncommutative Instantons}, Nucl.Phys. B621 (2002) 101,
hep-th/0108007.

\bibitem{KLY} K.-Y. Kim, B.-H. Lee and H. S. Yang, Phys.
Lett. B \textbf{523} (2001) 357 [hep-th/0109121].

\bibitem{CGOT} E.~Corrigan, P.~Goddard, H.~Osborn and S.~Templeton,
\textit{Zeta Function Regularization And Multi-Instanton
Determinants}, Nucl. Phys. B \textbf{159} (1979) 469.

\bibitem{Tian2} Y.~Tian and C.-J.~Zhu, {\it Remarks on the noncommutative ADHM construction}, Phys. Rev. D {\bf 67} (2003) 045016 [hep-th/0210163].

\end{thebibliography}
\end{document}